\documentclass[a4paper,preprint]{emulateapj}
\usepackage{amstext}
\usepackage{apjfonts}
\usepackage{amsmath}
\newcommand{\rb}[1]{\raisebox{1.5ex}[0pt]{#1}}

\begin{document}

\title{The Most Probable Cause for the High Gamma-Ray Polarization 
in GRB 021206}




\author{Jonathan Granot}
 
\affil{Institute for Advanced Study, Olden Lane, Princeton, 
NJ 08540; granot@ias.edu}

\begin{abstract}
  
  The exciting detection of a very high degree of linear polarization,
  $P=80\%\pm 20\%$, in the prompt $\gamma$-ray emission of the recent
  GRB 021206, provides strong evidence that synchrotron emission is
  the dominant radiation mechanism. Besides this immediate
  implication, there were also claims that this implies a magnetic
  field that is ordered on large scales within the ejecta, and must
  therefore be produced at the source, which in turn was used as an
  argument in favor magnetic fields playing an active role in the
  production of GRB jets. However, an alternative explanation was also
  suggested: a very narrow jet, of opening angle $\theta_j\sim
  1/\gamma$, where $\gamma\gtrsim 100$ is the Lorentz factor during
  the GRB, viewed slightly outside its edge, at $\theta_j<\theta_{\rm
    obs}\lesssim\theta_j+1/\gamma$. This explanation also works with a
  magnetic field that is generated in the internal shocks and does not
  originate at the source. We calculate the expected degree of
  polarization for these two scenarios, and find that it is
  significantly easier to produce $P\gtrsim 50\%$ with an ordered
  field. More specifically, we obtain $P\sim 43-61\%$ for an ordered
  transverse magnetic field, $B_{\rm ord}$, whereas a shock-produced
  field that is random but fully within the plane of the shock,
  $B_\perp$, can produce up to $P\lesssim 38-54\%$ for a single pulse
  in the GRB light curve, but the integrated emission over many pulses
  (as measured in GRB 021206) is expected to be a factor of $\sim 2$
  lower. A magnetic field normal to the shock front, $B_\parallel$,
  can produce $P\sim 35-62\%$ for the emission integrated over many
  pulses. However, polarization measurements from GRB afterglows
  suggest a more isotropic configuration for the shock-produced field
  that should reduce $P$ by a factor $\sim 2-3$.  Therefore, an
  ordered magnetic field, $B_{\rm ord}$, that originates at the
  source, can produce the observed polarization most naturally, while
  $B_\parallel$ is less likely, and $B_\perp$ is the least
  likely of the above.

\end{abstract}



\section{Introduction}
\label{introduction}

The recent detection of a very large linear polarization, $P=80\%\pm
20\%$, in the prompt $\gamma$-ray emission of GRB 021206 (Coburn \&
Boggs 2003, hereafter CB), establishes synchrotron emission as the
dominant radiation mechanism in the prompt GRB. As the prompt GRB is
believed to arise from internal shocks within a relativistic outflow
(Rees \& M\'esz\'aros 1994; Sari \& Piran 1997), it can provide
valuable information on the magnetic field structure in the ejecta,
and clues to the nature of the central source.  In a recent paper
(Granot \& K\"onigl 2003, hereafter GK), we suggested that ``the
radiation from the original ejecta, which includes the prompt GRB and
the emission from the reverse shock (the `optical flash' and `radio
flare') could potentially exhibit a high degree of polarization (up to
$\sim 60\%$) induced by an ordered transverse magnetic field advected
from the central source''.  This is perfectly consistent with the
polarization measured in GRB 021206. CB also attributed the
polarization in this GRB to an ordered magnetic field, and suggested
that this implies that magnetic fields drive the GRB explosion.  A
similar interpretation of this measurement has even been claimed to
favor Poynting dominated outflows in GRBs (Lyutikov, Periev \&
Blandford 2003).

However, Waxman (2003) suggested an alternative explanation: if the
GRB outflow is a uniform jet with sharp edges and an opening angle
$\theta_j\lesssim 1/\gamma$, then our line of sight is likely to be at
an angle $\theta_j<\theta_{\rm obs}\lesssim\theta_j+1/\gamma$ from the
jet axis. In this case we should see both a bright GRB (as much of the
radiation is still beamed toward us) and a large polarization (e.g.
Gruzinov 1999; Granot et al. 2002). This scenario does not require an
ordered field and also works for a magnetic field that is generated at
the internal shocks (Medvedev \& Loeb 1999).

There are therefore two feasible explanations for the large
polarization measured in GRB 021206, where only one of them requires a
magnetic field ordered on angular scales $\gtrsim 1/\gamma$. This
undermines the possible theoretical implications of an ordered
magnetic field in the GRB ejecta. In this Letter we critically examine
these two scenarios, and estimate their ability to explain the high
observed polarization.  In \S \ref{B_ord} we calculate the
polarization from an ordered magnetic field.  The maximal polarization
for a narrow jet with a shock-produced magnetic field is calculated in
\S \ref{narrow_jet}.  In \S \ref{GRB021206} we apply our results to
GRB 021206 and discuss the conclusions.

\section{An Ordered Magnetic Field}
\label{B_ord}

Here we calculate the linear polarization for synchrotron emission from
a thin spherical shell with an ordered transverse magnetic field, $B_{\rm
  ord}$, moving radially outward with $\gamma\gg 1$. We integrate over
the emission from the shell at a fixed radius and do not follow the
different photon arrival times from different angles $\theta$ from the
line of sight (l.o.s.).  This calculation is relevant to the prompt
GRB, the reverse shock (the `optical flash' and `radio flare') and the
afterglow, provided the magnetic field is ordered over an angle
$\gtrsim 1/\gamma$ around the l.o.s.

Following GK, the polarization position angle, measured from
$\hat{B}_{\rm ord}$, is given by
$\theta_p=\phi+\arctan(\frac{1-y}{1+y}\cot\phi)$ in the limit
$\gamma\gg 1$, where $y\equiv(\gamma\theta)^2$ and $\phi$ is the
azimuthal angle.  We have $I_\nu=I'_{\nu'}(\nu/\nu')^3$,
with\footnote{\label{epsilon} Here $\chi'$ is the angle between
  $\hat{n}'$ and $\hat{B}'$, which is also the pitch angle between the
  electron's velocity and $\hat{B}'$. For the optically thin part of
  the spectrum that is considered in this work, and as long as the
  electron energy distribution (taking into account electron cooling)
  is independent of the pitch angle $\chi'$ (which is most natural for
  a random field, and is also reasonable to expect, at least
  approximately, for an ordered field as well), we find
  $\epsilon=1+\alpha$.}
$I'_{\nu'}\propto(\nu')^{-\alpha}(\sin\chi')^\epsilon
\propto(\nu')^{-\alpha}[1-(\hat{n}'\cdot\hat{B}'_{\rm
  ord})^2]^{\epsilon/2}$, where $\nu/\nu'\approx 2\gamma/(1+y)$,
$1-(\hat{n}'\cdot\hat{B}'_{\rm ord})^2\approx
[(1-y)/(1+y)]^2\cos^2\phi+\sin^2\phi$ and $\hat{n}'$ is the direction
in the local frame of a photon that reaches the observer.  The Stokes
parameters are given by
\begin{equation}\label{Stokes}
\frac{\left\{\begin{matrix} U \cr Q \end{matrix}\right\}}
{IP_{\rm max}}=\frac{\int \frac{dy}{(1+y)^{a}}\int d\phi
\left[\left(\frac{1-y}{1+y}\right)^2\cos^2\phi
+\sin^2\phi\right]^{\epsilon/2}
\left\{\begin{matrix} \sin 2\theta_p \cr \cos 2\theta_p
 \end{matrix}\right\}}{\int\frac{dy}{(1+y)^{a}}\int d\phi
\left[\left(\frac{1-y}{1+y}\right)^2\cos^2\phi
+\sin^2\phi\right]^{\epsilon/2}}
\ ,
\end{equation}
where $a=3+\alpha$ for the instantaneous emission (relevant for the
afterglow) and $a=2+\alpha$ for the time integrated emission (relevant
for the prompt GRB when integrated over a time larger than the
duration of a single pulse, as in GRB 021206). For a uniform jet, the
limits of integration should include only regions within the jet.
This is important only if $\theta_{\rm obs}+1/\gamma\gtrsim\theta_j$,
which is expected to be rare for the prompt GRB, but usually occurs
during the afterglow.  When the edge of the jet is at $y>y_{\rm
  max}\gtrsim {\rm a\ few}$, the limits of integration may be taken as
$\int_0^{y_{\rm max}}dy\int_0^{2\pi}d\phi$. In this case $U=0$ and
$P_{\rm ord}=-Q/I=|Q|/I$. For internal shocks, each pulse in the GRB
light curve is from a collision between two shells. The emission near
the peak of the pulse is mainly from $\theta\lesssim
1/\gamma$ ($y\lesssim 1$), and may be approximated by taking $y_{\rm
  max}=1$. The emission from $y\gtrsim 1$ contributes mainly to the
tail of the pulse. If the latter is included in the temporal
integration used for measuring $P$, and is not below
the background, then we can take $y_{\rm max}\gg 1$ (the asymptotic
limit is reached at $y_{\rm max}\gtrsim\;$a few).

\begin{figure}
  \epsscale{0.85} 
\plotone{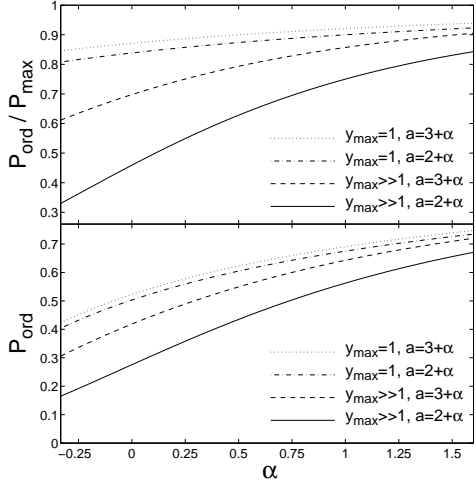} 
\figcaption[]{\label{fig1}The
    polarization $P_{\rm ord}$ (lower panel) and $P_{\rm ord}/P_{\rm
      max}$ (upper panel) of synchrotron emission from an ordered
    transverse magnetic field, as a function of the spectral index
    $\alpha$, for $\epsilon=1+\alpha$, calculated using Eq.
    (\ref{Stokes}).}
\end{figure}

\begin{deluxetable}{cccccc}
\tablewidth{300pt}
\tablecaption{Parameter Values for Different PLSs of the Spectrum
\label{table1}}
\tablehead{\colhead{PLS} & \colhead{$\alpha$} & \colhead{$P_{\rm max}$} & 
\colhead{$P_{\rm ord}(y_{\rm max}=1)$} &
\colhead{$P_{\rm ord}(y_{\rm max}\gg1)$}}
  \startdata
  D,\,E & -1/3    & 1/2  & 0.404 (0.423)   &  0.165 (0.306)   \\
  F & 1/2     & 9/13 & 0.605 (0.623)   &  0.435 (0.549)   \\
  &         &      & 0.605--0.675    &  0.435--0.563          \\
  \rb{G} & \rb{\large $\frac{p-1}{2}$} &
  \rb{\large $\frac{p+1}{p+7/3}$} & (0.623--0.691) & (0.549--0.643) \\
  &  &  & 0.675--0.726 &  0.563--0.656          \\
  \rb{H} & \rb{$p/2$} &
  \rb{\large $\frac{p+2}{p+10/3}$} & (0.691--0.739) & (0.643--0.709) \\
  \hline \enddata \tablecomments{Parameter values for different power
    law segments (PLSs) of the spectrum (that are labeled as in Granot
    \& Sari 2002). Numerical values in PLSs G and H are for an
    electron index $2<p<3$. The values of $P_{\rm ord}$ without (with)
    parentheses are for $a=2+\alpha$ ($a=3+\alpha$) which are
    appropriate for the prompt GRB (afterglow). (see discussion below
    Eq. \ref{Stokes}).}
\end{deluxetable}

Figure \ref{fig1} shows $P_{\rm ord}/P_{\rm max}$ and\footnote{Here
  $P_{\rm max}=(\alpha+1)/(\alpha+5/3)=(p_{\rm eff}+1)/(p_{\rm
    eff}+7/3)$, where it is useful to define $p_{\rm eff}\equiv
  2\alpha+1$. For optically thin synchrotron emission, $\alpha\geq
  -1/3$, and hence $P_{\rm max}\geq 1/2$. This lower limit on $P_{\rm
    max}$ arises since $P=1/2$ is simply the low frequency
  ($\nu\ll\nu_{\rm syn}$) polarization of the synchrotron emission
  from each electron, and therefore $P_{\rm max}=1/2$ in PLSs D,E (see
  Table 1). For PLSs F, G and H, $P_{\rm max}$ is determined by
  $p_{\rm eff}$ ($=2$, $p$, and $p+1$, respectively), where for these
  PLSs, $p_{\rm eff}$ is the effective power law index of the electron
  distribution.}  $P_{\rm ord}$ as a function of $\alpha$ for
$\epsilon=1+\alpha$ (e.g.  footnote \ref{epsilon}), and Table 1
summarizes the results for the relevant (optically thin) power law
segments (PLSs) of the spectrum.\footnote{\label{formula} The most
  relevant case for GRB 021206 is $y_{\rm max}\gg 1$ and $a=2+\alpha$,
  for which the approximate formula $P_{\rm ord}(\alpha)=0.016\alpha^4
  -0.052\alpha^3-0.013\alpha^2+0.335\alpha+0.276$, provides a relative
  accuracy of better than $0.25\%$ for $-1/3\leq\alpha\leq
  3/2$.}

\section{A Very Narrow Jet Viewed  from Just Outside its edge}
\label{narrow_jet}

In this section we calculate the polarization from a narrow jet, of
opening angle $\theta_j\sim 1/\gamma$, viewed at an angle
$\theta_j\lesssim\theta_{\rm obs}\lesssim\theta_j+1/\gamma$ from its
axis.\footnote{It is assumed here that the jet has sharp
  edges, i.e. the emissivity drops sharply over an
  angular interval $\Delta\theta\ll 1/\gamma$ around
  $\theta=\theta_j$. A smoother edge, $\Delta\theta\gtrsim 1/\gamma$,
  would considerably reduce the polarization.} In contrast to \S
\ref{B_ord}, here the magnetic field is assumed to be produced at the
shock itself, and therefore has symmetry around the direction normal
to the shock, $\hat{n}_{\rm sh}$. Since the more isotropic the
magnetic field configuration behind the shock, the lower the resulting
polarization, we consider two extreme cases where the field is most
anisotropic: 1.  a random field that lies strictly within the plane of
the shock ($B=B_\perp$, $P=P_\perp$), 2. a completely ordered field in
the direction of $\hat{n}_{\rm sh}$ ($B=B_\parallel$,
$P=P_\parallel$).

\begin{figure}
\epsscale{1}
\plotone{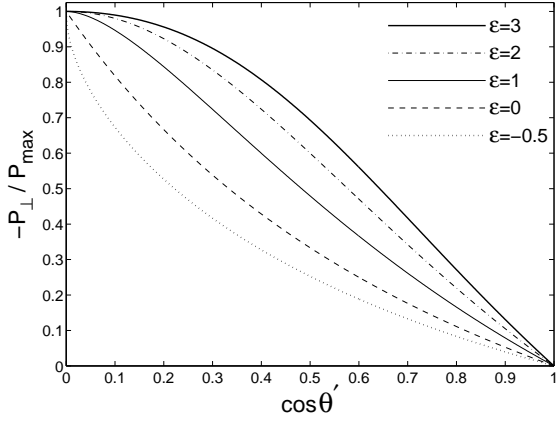}
\figcaption[]{\label{fig2}The local polarization, 
$|P_\perp(\theta')|=-P_\perp(\theta')$, 
normalized by $P_{\rm max}$, for a magnetic 
field that is fully tangled within a plane, and emission at an 
angle $\theta'$ from the normal to the plane, 
for $\epsilon=-0.5,0,1,2,3$.}
\end{figure}

Following Ghisellini \& Lazzati (1999), we generalize their 
formula so that it would hold for $\theta_{\rm obs}>\theta_j$, 
\begin{equation}\label{P_perp_local}
P=\frac{\frac{1}{2\pi}
\int_{|\theta_j-\theta_{\rm obs}|}^{\theta_j+\theta_{\rm obs}}
\theta d\theta I_{\nu}(\theta)P(\theta)
\sin[2\psi_1(\theta)]}{\Theta(\theta_j-\theta_{\rm obs})
\int_0^{\theta_j-\theta_{\rm obs}}\theta d\theta 
I_{\nu}(\theta)
+\int_{|\theta_j-\theta_{\rm obs}|}^{\theta_j+\theta_{\rm obs}}
\theta d\theta\frac{\pi-\psi_1(\theta)}{\pi}I_{\nu}(\theta)}\ ,
\end{equation}
where $\cos\psi_1= (\theta_j^2-\theta_{\rm
  obs}^2-\theta^2)/2\theta_{\rm obs}\theta$ and $\Theta(x)$ is the
Heaviside step function.  For $B_\parallel$ we simply have
$P_\parallel(\theta)=P_{\rm max}$ and
$\hat{n}'\cdot\hat{B}'=\hat{n}'\cdot\hat{r}=
\cos\theta'\approx\frac{1-y}{1+y}$, so that $I_\nu\propto
y^{\epsilon/2}/(1+y)^{3+\alpha+\epsilon}$.  However, for $B_\perp$ we
must average over the possible field orientations within the plane of
the shock:\footnote{Here $P<0$ ($P>0$) implies $\hat{P}$ along
  (perpendicular to) the plane containing $\hat{n}_{\rm sh}$ and
  $\hat{n}'$.}
\begin{equation}\label{PPmax_par}
\frac{P_\perp(y)}{P_{\rm max}}=\frac{\int_0^{\pi}d\phi
\left[1-\frac{4y\cos^2\phi}{(1+y)^2}\right]^{(\epsilon-2)/2}
\left[\left(\frac{1-y}{1+y}\right)^2\cos^2\phi-\sin^2\phi\right]}
{\int_0^{\pi}d\phi
\left[1-\frac{4y\cos^2\phi}{(1+y)^2}\right]^{\epsilon/2}}\;,
\end{equation}
and $I_\nu\propto(1+y)^{-3-\alpha}$ times the denominator of Eq.
(\ref{PPmax_par}). For $\epsilon=2$ and $0$, $P_\perp(\theta')/P_{\rm
  max}=\frac{-2y}{1+y^2}= \frac{-\sin^2\theta'}{1+\cos^2\theta'}$ and
$-\min(y,1/y)$, respectively.  Fig. \ref{fig2} shows
$-P_\perp(\theta')/P_{\rm max}$ for several values of $\epsilon$.  A
larger $\epsilon$ implies a larger $|P(\theta')|$, as it suppresses
$I_\nu$ at $(\hat{n}'\cdot\hat{B}')^2\approx 1$ where there is a
positive contribution to $P_\perp(\theta')$.  The global polarization
from the whole jet is given by\footnote{Here, $P<0$ ($P>0$) means
  $\hat{P}$ along (perpendicular to) the direction from our l.o.s. to
  the jet axis.}
\begin{eqnarray}\label{P_par}
\frac{P_\parallel}{P_{\rm max}}&=&\frac{\frac{1}{2\pi}\int_{y_1}^{y_2}
\frac{y^{\epsilon/2}dy}{(1+y)^{a+\epsilon}}
\sin\left[2\psi_1(y)\right]}
{\Theta(1-q)\int_0^{y_1}\frac{y^{\epsilon/2}dy}{(1+y)^{a+\epsilon}}
+\int_{y_1}^{y_2}\frac{y^{\epsilon/2}dy}{(1+y)^{a+\epsilon}}
\frac{\left[\pi-\psi_1(y)\right]}{\pi}}\ ,
\\ \label{P_perp}
\frac{P_\perp}{P_{\rm max}}&=&\frac{\frac{1}{2\pi}\int_{y_1}^{y_2}
\frac{dy\sin\left[2\psi_1(y)\right]}{(1+y)^{a}}g(y,\epsilon)}
{\Theta(1-q)\int_0^{y_1}\frac{f(y,\epsilon)dy}{(1+y)^{a}}
+\int_{y_1}^{y_2}\frac{f(y,\epsilon)dy}{(1+y)^{a}}
\frac{\left[\pi-\psi_1(y)\right]}{\pi}}\ ,
\\ \label{f_y_eps}
f(y,\epsilon)&=&\int_0^\pi d\phi
\left[1-\frac{4y\cos^2\phi}{(1+y)^2}\right]^{\epsilon/2}\ ,
\\ \label{g_y_eps}
g(y,\epsilon)&=&\int_0^\pi d\phi\,
\frac{(1-y)^2(1+y)^{-2}\cos^2\phi-\sin^2\phi}
{\left[1-4y(1+y)^{-2}\cos^2\phi\right]^{(2-\epsilon)/2}}\ ,
\end{eqnarray}
where $\cos\psi_1= [(1-q^2)y_j-y]/(2q\sqrt{y_jy})$,
$q\equiv\theta_{\rm obs}/\theta_j$, $y_j=(\gamma\theta_j)^2$,
$y_1=(1-q)^2y_j$, $y_2=(1+q)^2y_j$.

\begin{figure}
\epsscale{0.85}
\plotone{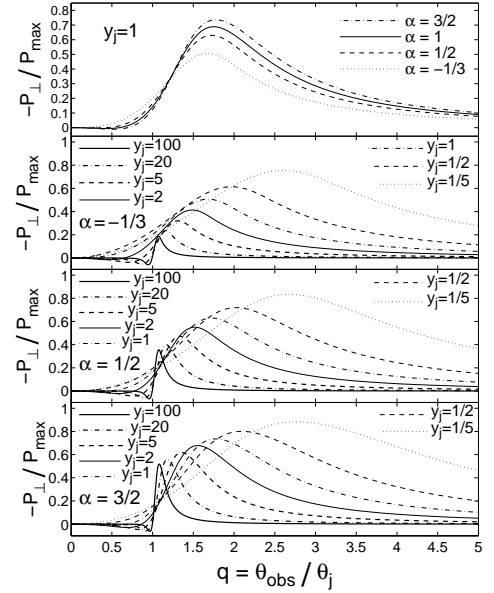}
\figcaption[]{\label{fig3}The polarization $-P_\perp(q)/P_{\rm max}$
  for several values of $\alpha$ and $y_j$, calculated using Eq.
  (\ref{P_perp}) with $a=2+\alpha$ and $\epsilon=1+\alpha$.}
\end{figure}

\begin{figure}
\epsscale{0.85}
\plotone{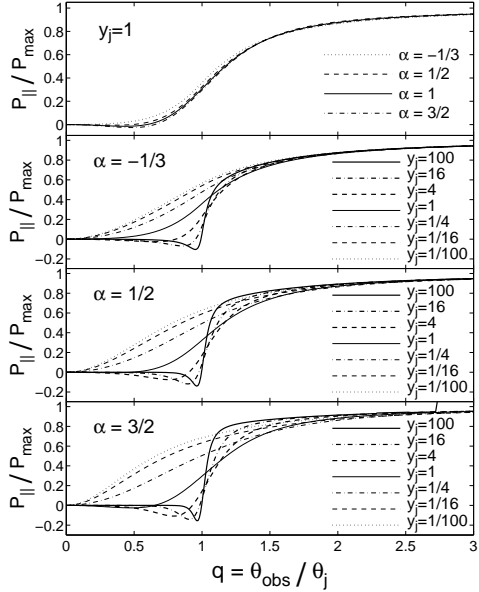}
\figcaption[]{\label{fig4}The same as Fig. \ref{fig3} but for
  $P_\parallel$ using Eq. (\ref{P_par}).}
\end{figure}

Figures \ref{fig3} and \ref{fig4} show $P_\perp(q)$ and
$P_\parallel(q)$, respectively, for several values of $\alpha$ and
$y_j$, using the relation $\epsilon=1+\alpha$.  For $q<1$,
$|P_\perp|/P_{\rm max}\lesssim 0.2$, while $|P_\perp|$ rises sharply
above $q=1$ (the larger $y_j$, the sharper the rise), and peaks at
$q\sim 1+1/\sqrt{y_j}$ ($q\approx 1.7-1.8$ for $y_j=1$), which is $q$
just above $1$ for $y_j\gg 1$, but at $q\sim 1/\sqrt{y_j}\gg 1$ for
$y_j\ll 1$. The width of the peak is $\sim 1/\sqrt{y_j}$, so that the
peak is wider (as well as higher) for smaller $y_j$.  At larger values
of $q$, $|P_\perp|$ decreases since for $\theta_{\rm
  obs}\gtrsim(2-3)\max(\theta_j,1/\gamma)$ [i.e.
$q\gtrsim(2-3)\max(1,1/\sqrt{y_j})$], the jet may be approximated as a
point source, and as $q$ increases, the emission in the local frame is
almost straight backward (i.e. $\hat{n}'$ approaches $-\hat{n}_{\rm
  sh}$ and $\theta'$ approaches $\pi$), thus suppressing
$P_\perp(\theta')$ (see Fig. \ref{fig2}).  However, in sharp contrast
with $B_\perp$, for $B_\parallel$ even if $\theta'$ is only slightly
different from $\pi$, still $P_\parallel(\theta')=P_{\rm max}$, and
$P_\parallel$ approaches $P_{\rm max}$ for $q\gtrsim 2$. The
transition between $P_\parallel(q\gtrsim 2)\approx P_{\rm max}$ and
$P_\parallel(q=0)=0$ is very gradual for $y_j\ll 1$ and very sharp for
$y_j\gg 1$ (for which the transition occurs at $|q-1|\lesssim
y_j^{-1/2}\ll 1$).  For $y_j>1$, $P_\parallel(q<1)/P_{\rm max}\lesssim
0.3$, which is a little higher than $|P_\perp|$, while for $y_j<1$,
$P_\parallel(q<1)/P_{\rm max}\lesssim 0.6$.  For $y_j\gg 1$
($\theta_j\gg 1/\gamma$) the edge of the jet is hardly visible from
the interior of the jet and $P(q<1-1/\sqrt{y_j})\approx 0$ (for both
$B_\perp$ and $B_\parallel$).

The above expressions for $P_{\perp}(q,y_j,\epsilon,a)$ or
$P_{\parallel}(q,y_j,\epsilon,a)$ can produce afterglow polarization
light curves, by using $a=3+\alpha$, $\epsilon=1+\alpha$ and $P_{\rm
  max}=\frac{\alpha+1}{\alpha+5/3}$, and adding a model for the time
evolution of $\gamma(t)$ and $\theta_j(t)$, which determine
$q(t)=\theta_{\rm obs}/\theta_j(t)$ and
$y_j(t)=[\gamma(t)\theta_j(t)]^2$.  One simple model is to assume
$q(t<t_j),y_j(t>t_j)={\rm const}$, where $t_j$ is the jet break
time.\footnote{Ghisselini \& Lazzati (1999) simply assumed
  $\theta_j,\alpha={\rm const}$, and implicitly assumed $\epsilon=2$
  since they used $P(\theta')/P_{\rm
    max}=\sin^2\theta'/(1+\cos^2\theta')$.  However, they did not take
  into account the fact that
  $I'_{\nu'}\propto\langle(\sin\chi')^\epsilon\rangle\propto
  \langle[1-(\hat{B}'\cdot\hat{n}')^2]^{\epsilon/2}\rangle$, which
  effects the polarization light curves.} Note that at a fixed
observed time, $P$ remains constant within each PLS, but changes
across spectral breaks.

\section{Application to GRB 021206 and Discussion} 
\label{GRB021206}

In the prompt GRB the spectral index is usually in the range
$1/2\lesssim\alpha\lesssim 5/4$, for which the time integrated
polarization ($a=2+\alpha$, $y_{\rm max}\gg 1$) from an ordered
transverse magnetic field ($B_{\rm ord}$) is $P_{\rm ord}\sim 43-61\%$
(e.g. Table 1, Fig. \ref{fig1}, footnote \ref{formula}).  This is
reasonably consistent with the value of $P=80\%\pm 20\%$ that was
measured for GRB 021206 (CB). Furthermore, this requires a magnetic
field that is ordered over angles $\gtrsim 1/\gamma$ which can still
be $\ll\theta_j$.

We now turn to the narrow jet scenario (Waxman 2003).  For
$\theta_{\rm obs}>\theta_j+1/\gamma$ the observed flux from the GRB
drops considerably. Therefore, a bright GRB like 021206 requires
$q\lesssim 1+1/\sqrt{y_j}$.  As it is hard to collimate a jet to
$\theta_j<1/\gamma$, it is reasonable to assume $y_j\gtrsim 1$ and
therefore $q\lesssim 2$.  Furthermore, for $y_j\gg 1$ that is usually
inferred from afterglow observations (Panaitescu \& Kumar 2002), the
peak of the polarization is at $q\sim 1+1/\sqrt{y_j}\sim 1$ and has a
width $\Delta q\sim 1/\sqrt{y_j}\ll 1$ which covers a fraction $\sim
1/\sqrt{y_j}\ll 1$ of the solid angle from which the GRB is beamed
toward us, and therefore a high polarization is very unlikely.
Hence, we require $y_j\lesssim 2$.  The fact that GRB 021206 was
extraordinarily bright, together with the correlation found by Frail
et al. (2001), might suggest a very narrow jet, so that $y_j\lesssim
2$ is not so far fetched (Waxman 2003).  Altogether we expect
$1\lesssim y_j\lesssim 2$ and $1\lesssim q\lesssim 2$.

In this parameter range, and for $1/2\lesssim\alpha\lesssim 5/4$,
$P_\perp$ peaks at $P_{\perp,{\rm max}}\sim(0.55-0.7)P_{\rm max}\sim
38-54\%$.  However, the Lorentz factor of the shocked fluid in the
internal shocks is expected to vary with $\Delta\gamma\sim\gamma$
between different shell collisions within the same GRB.\footnote{If
  $B_{\rm ord}$ is ordered on angles $\gtrsim 1/\gamma_{\rm min}$,
  which are still $\lesssim 0.01$, and $\hat{B}_{\rm ord}$ does not
  change significantly (i.e. by $\lesssim 0.5\;$radians) between the
  different shells, then this should not effect $P_{\rm ord}$
  significantly; $P_\parallel$ should also not be strongly effected.}
This implies a reasonably large variation in $y_j\propto\gamma^2$,
while $q={\rm const}$, so that our l.o.s. will not be near the peak of
$P_\perp$ for all the pulses in the GRB light curve. Furthermore, the
observed flux at $q\sim 1+1/\sqrt{y_j}$, where $P_{\perp}$ peaks, is
smaller than near the edge of the jet ($q\approx 1$), due to
relativistic beaming effects, so that the brightest pulses would tend
to be relatively weakly polarized, thus further reducing the average
polarization over the whole GRB.  Therefore, while for a single pulse
in the GRB light curve $P_\perp$ can approach $P_{\perp,{\rm max}}$,
the average over many pulses (as in GRB 021206) will be
$P_\perp\lesssim P_{\perp,{\rm max}}/2\sim 19-27\%$.
 
For $B_\parallel$ we find $P_\parallel\sim(0.3-0.9)P_{\rm max}\sim
20-70\%$ for a single pulse, and expect
$P_\parallel\sim(0.5-0.8)P_{\rm max}\sim 35-62\%$ for the average over
many pulses, which is consistent with the value measured for GRB
021206. In fact, $B_\parallel$ is an ordered magnetic field, just that
unlike $B_{\rm ord}$ which was considered in \S \ref{B_ord}, it can in
principle be generated at the shock itself, as $\hat{n}_{\rm sh}$ is a
preferred direction that is determined locally by the shock front.
Current models for the production of magnetic fields at collisionless
relativistic shocks (Medvedev \& Loeb 1999) suggest $B=B_\perp$ rather
than $B_\parallel$. However the amplification mechanism of the
magnetic field and its configuration in relativistic shocks is still
largely an open question, so that it is hard to rule out $B\approx
B_\parallel$ on purely theoretical grounds.  Nevertheless, it is
important to keep in mind that we considered two extreme cases for the
magnetic field configuration behind the shock, in which it is most
anisotropic. The relatively low values of $P\lesssim 3\%$ measured in
GRB afterglows\footnote{except for a possible sharp spike with
  $P\approx 10\%$ in the polarization light curve of GRB 020405
  (Bersier et al.  2003).}, compared to the expected values of
$P\lesssim 20\%$ (Sari 1999; Ghisellini \& Lazzati 1999; GK), suggest
that the magnetic field created behind relativistic shocks is more
isotropic than the extreme cases we considered, implying $P$ values
lower by a factor of $\sim 2-3$ (e.g.  GK).  Therefore, although
$P_\parallel\sim 35-62\%$, a more isotropic magnetic field
configuration that is suggested by afterglow observations would
imply\footnote{The same argument should reduce $P_\perp\sim 19-27\%$
  to $P\sim 7-13\%$, making it even harder to reconcile with the value
  measured in GRB 021206.}  $P\sim 15-30\%$.

We therefore conclude that $P\gtrsim 50\%$ is most naturally produced
by an ordered magnetic field that is carried out with the ejecta from
the central source (as was recently proposed by GK).  This is
therefore the most likely explanation for the value of $P=80\%\pm 20\%$
(CB) measured in GRB 021206. A magnetic field that is generated at the
shock itself is less likely to produce a sufficiently large
polarization. However, if either 1) the systematic uncertainty in the
quoted value was for some reason underestimated, and $P\lesssim
20-30\%$ is acceptable, or 2) the internal shocks for some reason
produce a magnetic field much more anisotropic than in the afterglow
shock, then $P_\parallel$ may still be a viable option. Both points are
required in order for $P_\perp$ to work well.

\acknowledgments I thank Davide Lazzati, Arieh K\"onigl, Ehud Nakar
and Eli Waxman for useful discussions.  This research was supported by
the Institute for Advanced Study, funds for natural sciences.

\end{document}